\documentclass[onecolumn,10pt,showpacs,
amsmath,amssymb]{revtex4}
\usepackage{graphicx}
\usepackage{hyperref}
\usepackage{amssymb, amsmath}
\usepackage{epsf}

\def\la{\; \raise0.3ex\hbox{$<$\kern-0.75em\raise-1.1ex\hbox{$\sim$}}\;}
\def\ga{\;  \raise0.3ex\hbox{$>$\kern-0.75em\raise-1.1ex\hbox{$\sim$}}\;}

\def\pFn{p_{\raise-0.3ex\hbox{{\scriptsize F$\!$\raise-0.03ex\hbox{\rm n}}}}
}  

\def\pFa{p_{\raise-0.3ex\hbox{{\scriptsize F$\!$\raise-0.03ex\hbox{$i$}}}}
}  

\def\pFas{p_{\raise-0.3ex\hbox{{\scriptsize F$\!$\raise-0.03ex\hbox{$k$}}}}
}  

\def\pFb{p_{\raise-0.3ex\hbox{{\scriptsize F$\!$\raise-0.03ex\hbox{$\beta$}}}}
}  

\def\vFa{v_{\raise-0.3ex\hbox{{\scriptsize F$\!$\raise-0.03ex\hbox{$i$}}}}
}  
\def\pFp{p_{\raise-0.3ex\hbox{{\scriptsize F$\!$\raise-0.03ex\hbox{\rm p}}}}
}  
\def\pFe{p_{\raise-0.3ex\hbox{{\scriptsize F$\!$\raise-0.03ex\hbox{\rm e}}}}
}  
\def\pFmu{p_{\raise-0.3ex\hbox{{\scriptsize F$\!$\raise-0.03ex\hbox{\rm
$\mu$}}}} }  
\def\m@th{\mathsurround=0pt }
\def\eqalign#1{\null\,\vcenter{\openup1\jot \m@th
   \ialign{\strut$\displaystyle{##}$&$\displaystyle{{}##}$\hfil
   \crcr#1\crcr}}\,}

\newcommand{\dd}{\mbox{d}}                     

\begin{document}

\title{Temperature effects in pulsating superfluid neutron stars}
%
\author{Elena M. Kantor$^{1,2}$ and Mikhail E. Gusakov$^{1}$}
\affiliation{
$^1$Ioffe Physical Technical Institute,
Polytekhnicheskaya 26, 194021 St.-Petersburg, Russia
\\
$^2$St.-Petersburg State Polytechnical University,
Polytekhnicheskaya 29, 195251 St.-Petersburg, Russia
}

\date{}
%


\begin{abstract}
We study the effects of finite stellar temperatures on the oscillations
of superfluid neutron stars.
The importance of these effects is illustrated with 
a simple example of a radially pulsating general relativistic star. 
Two main effects are taken into account: 
(i) temperature dependence of the entrainment matrix 
and (ii) the variation 
of the size of superfluid region with temperature.
Four models are considered, 
which include either one or both of these two effects. 
Pulsation 
spectra are calculated for these models, 
and asymptotes for eigenfrequencies 
at temperatures close to
critical temperature of neutron superfluidity, 
are derived. 
It is demonstrated that models that 
allow for 
the temperature effect (ii) but disregard the effect (i),
yield 
unrealistic results. 
Eigenfunctions for the normal- and superfluid-type 
pulsations are analyzed. 
It is shown that superfluid pulsation modes practically 
do not appear at the neutron-star surface and, 
therefore, can hardly be observed by measuring 
the modulation of the electromagnetic radiation from the star.
The e-folding times for damping of pulsations
due to the shear viscosity 
and nonequilibrium modified Urca processes
are calculated and their asymptotes 
at temperatures close to the neutron critical temperature, are obtained. 
It is demonstrated that superfluid pulsation modes
are damped by 1--3 orders of magnitude faster than normal modes.
\end{abstract}

\maketitle

\section{Introduction}
\label{1}

High-frequency oscillations of the electromagnetic radiation, 
interpreted as resulting from neutron-star oscillations, 
have already been observed during giant flares \cite{sw2005,israel2005}.
Also, detectors are being designed, which, 
according to preliminary estimates, 
will be able to detect gravitational radiation from 
oscillating neutron stars (NSs) \cite{ak01,andersson03,owen10,ligo07}. 
To interpret the existing and future observations, 
it is necessary to have a well-developed theory of neutron-star oscillations.
Our study is devoted to 
the superfluid sector of 
this theory. 
To be more concrete, the goal of the present paper 
is to demonstrate,
using radial pulsations as an example,
that the effects of finite temperatures 
strongly affect 
dynamics of NSs.

Neutron stars are relativistic objects 
having supranuclear densities 
and strongly curving spacetime 
around them.
Therefore, accurate calculations of their properties 
should be made in the frame of the 
general relativity theory.
The situation is additionally complicated by the fact that baryons
in the neutron-star cores can be in superfluid state \cite{yls99,ls01,yp04}. 
Thus, several velocity fields can co-exist in matter without dissipation, 
which makes hydrodynamic equations 
substantially more complicated 
\cite{ll87,khalatnikov89,putterman74,gusakov07,ga06,lm94,cll99,ac01a}.
Epstein \cite{epstein88} was the first who considered 
sound waves in the superfluid matter of NSs. 
Global pulsations of superfluid Newtonian stars were first studied 
by Lindblom and Mendell \cite{lm94}.
Since then, 
a growing number of papers 
devoted to oscillations of superfluid NSs
have been published 
(see, e.g., \cite{ac01b,ga06,gusakov07,lee95,ac01a,
acl02,pca04,yl03a,yl03b,cll99,lm00,sac08,lac08} and references therein).
To make the problem tractable, 
most of these authors
employed the nonrelativistic hydrodynamics in their calculations.
Pulsations of general relativistic superfluid NSs were analyzed only 
in Refs.\ \cite{cll99,ac01b,acl02,yl03b,ga06,lac08} 
under a number of simplified assumptions. 
In particular, until recently it was customary to use 
the zero-temperature limit of superfluid hydrodynamics. 
However, as it was first shown in Ref.\ \cite{ga06} (hereafter GA06),
pulsation spectra of superfluid NSs 
can be very sensitive to variation of temperature.

The authors of GA06 
studied oscillations of superfluid NSs 
using the finite-temperature version of relativistic hydrodynamics for superfluid mixtures, 
described in detail in Refs.\ \cite{ga06,gusakov07,gk08,kg09}. 
For simplicity, only radial pulsations were analyzed 
and the simplest $npe$-matter 
composition of the stellar core was adopted, 
including neutrons ($n$), protons ($p$), and electrons ($e$). 
Neutrons and protons in the core were allowed to be superfluid, 
while free neutrons in the crust were treated as normal (nonsuperfluid). 
Owing to the fact that the relativistic entrainment matrix 
(analog of the superfluid density \cite{khalatnikov89} for mixtures) 
is a steep function of temperature \cite{gkh09b}, 
the spectrum obtained in GA06 was strongly temperature-dependent. 
The authors of GA06 neglected another temperature effect, 
the decrease in the size of the superfluid region with increasing temperature. 
It should be noted that this effect has been considered 
in several papers \cite{lac08,hap09,ha10}; 
however, the authors of these papers disregarded 
the temperature dependence of the entrainment matrix 
(that is, they used the zero-temperature limit of superfluid hydrodynamics). 

In the present study we, for the first time, 
analyze the influence of both of  
these temperature effects on the pulsation spectra.
For that we consider the more realistic profiles of
the neutron critical temperature $T_{cn}(r)$ than that
employed in GA06, where it was assumed that the red-shifted $T_{cn}$ 
is
constant throughout the core. 
We also compare the eigenfunctions for normal and superfluid pulsation modes, 
and calculate their characteristic damping times. 
The asymptotes for the eigenfrequencies and for e-folding times 
at temperatures close to $T_{cn}$ will also be examined.

The paper is organized as follows. 
In Sec.\ II we briefly discuss 
the superfluid hydrodynamics for 
mixtures 
and write down 
a system of equations describing radial oscillations 
of general relativistic 
superfluid NSs. 
In Sec.\ III, we present the pulsation spectra 
calculated for four different neutron-star models 
that
allow for
one or both of the temperature effects 
(the temperature dependence of the entrainment matrix 
and/or the variation of the size of the superfluid region with temperature). 
In Sec.\ IV, the approximate approach of Refs.\ \cite{gk10a,gk10b} 
is used
to obtain the asymptotes for pulsation eigenfrequencies 
at temperatures close to $T_{cn}$.
Sec.\ V is devoted to analysis of the eigenfunctions of radial pulsations.
Finally, in Sec.\ VI we study damping of radial pulsations 
due to the shear viscosity and nonequilibrium modified Urca processes. 
Section VII presents the summary.

In the following, unless otherwise stated,
the system of units used is one in which 
the Boltzmann constant $k_B$, 
the Planck constant $\hbar$, 
and the speed of light $c$ equal unity:
$k_B=\hbar=c=1$.

\section{Basic equations}
\label{2}

\subsection{Superfluid hydrodynamics}

Following GA06, we consider the simplest $npe$-matter composition 
of the neutron-star cores.
Because both protons and neutrons can be in the superfluid state, 
one has to use the relativistic hydrodynamics of superfluid mixtures 
to study the oscillations of NSs 
(see, e.g., \cite{ga06,gusakov07,gk08,kg09}).
The main distinctive feature of this hydrodynamics 
is the presence of several velocity fields in the mixture. 
In our case, these are the four-velocity $u^{\mu}$ 
of the `normal' component of matter 
(electrons and Bogoliubov excitations of neutrons and protons) 
as well as the four-velocities of superfluid neutrons $v_{s(n)}^{\mu}$ 
and superfluid protons $v_{s(p)}^{\mu}$. 
Below we briefly discuss the basic equations 
describing nondissipative motion 
of superfluid mixtures 
(for more detail, see \cite{ga06,gusakov07,gk08,kg09}). 

The continuity equation and 
the energy-momentum conservation law have the form
\begin{eqnarray}
j^{\mu}_{ (l) ; \, \mu} &=& 0, \quad 
j^{\mu}_{(i)} = n_i u^{\mu} + Y_{ik} w^{\mu}_{(k)}, \quad
j^{\mu}_{({ e})} = n_{ e} u^{\mu},
\label{currents}\\
T^{\mu \nu}_{; \,\mu} &=& 0, \quad  
T^{\mu \nu} = (P+\varepsilon) \, u^{\mu} u^{\nu} + P g^{\mu \nu} 
+ Y_{ik} \left( w^{\mu}_{(i)} w^{\nu}_{(k)} + \mu_i \, w^{\mu}_{(k)} u^{\nu} 
+ \mu_k \, w^{\nu}_{(i)} u^{\mu} \right). \,\,\,\,\,
\label{Tmunu} 
\end{eqnarray} 
Here and below the subscripts $i$ and $k$ refer to nucleons: $i$, $k = n$, $p$; 
the subscript $l$ runs over all particle species, $l=n$, $p$, $e$. 
Unless otherwise stated the summation is assumed over the repeated nucleon indices 
$i$, $k$ and over the spacetime indices $\mu$, $\nu$, and $\alpha$. 
In Eqs.\ (\ref{currents}) and (\ref{Tmunu}), 
$j^{\mu}_{(l)}$ is the particle four-current for species $l$; 
$T^{\mu \nu}$ is the energy-momentum tensor; 
$g^{\mu \nu}$ is the metric tensor; 
$n_l$ and $\mu_l$ are the number density 
and chemical potential of particle species $l$, respectively; 
$\varepsilon$ and $P$ are the energy density and pressure, respectively. 
The four-vector $w^{\mu}_{(i)}$ 
is related to the superfluid four-velocity 
$v^{\mu}_{s(i)}$ by equality: 
$w^{\mu}_{(i)} = \mu_i(v^{\mu}_{s(i)}-u^{\mu})$. 
In what follows, all the thermodynamic quantities 
(e.g., $n_l$ and $\varepsilon$) 
are defined (measured) in the reference frame 
comoving with the normal component of the fluid 
[in this frame $u^{\mu}=(1,0,0,0)$; 
notice that $u^{\mu}$ 
is normalized in such a way that $u_{\mu} u^{\mu}= -1$]. 
This imposes an additional constraint 
on the four-vector $w^{\mu}_{(i)}$ \cite{ga06, gusakov07},
\begin{equation}
u_{\mu} w^{\mu}_{(i)} =0.
\label{uw}
\end{equation}
Using this constraint and Eqs.\ (\ref{currents}) and (\ref{Tmunu})
one immediately finds that, in a relativistically invariant form,
the particle number densities are defined as 
$n_l=-u_{\mu}\,  j^{\mu}_{(l)}$ \cite{snoska1}, 
while the energy density is 
$\varepsilon = u_{\mu} u_{\nu} \, T^{\mu \nu}$.

The matrix $Y_{ik}$ in Eqs.\ (\ref{currents}) and (\ref{Tmunu}), 
named the relativistic entrainment matrix, 
is a generalization of the concept 
of superfluid density \cite{khalatnikov89} 
to the case of relativistic mixtures. 
In the nonrelativistic theory, 
a similar matrix was first considered 
by Andreev and Bashkin \cite{ab75}. 
The matrix $Y_{ik}$ is expressed in terms 
of the Landau parameters $F_1^{ik}$ 
of asymmetric nuclear matter 
and universal functions of temperature, 
$\Phi_i$, as described in \cite{gkh09b}. 
%
Generally, 
it can be presented as 
a function of density $\rho$
and the combinations $T/T_{cn}$ and $T/T_{cp}$:
$Y_{ik}=Y_{ik}(\rho, T/T_{cn}, T/T_{cp})$,
where $T_{cp}$ is the proton critical temperature.
%

For illustration, in Fig.\ 1 we plot the
elements $Y_{ik}$ as functions of temperature $T$
for the fixed baryon number density $n_b=3 n_0 = 0.48$ fm$^{-3}$.
The matrix elements are normalized 
to the constant 
$Y = 3n_0/\mu_{ n}(3n_0) \approx 2.60\times 10^{41}\, 
\rm erg^{-1}\,cm^{-3}$, 
where $n_0=0.16 \,\rm fm^{-3}$ 
is the density in atomic nuclei 
and $\mu_{ n}(3n_0)$ is the neutron chemical potential. 
Here and below, in all our calculations we use
the parameterization \cite{hh99} 
of equation of state (EOS) of Akmal et al. \cite{apr98} (hereafter APR).
The vertical dot-dashed lines indicate 
the critical temperatures for neutrons, 
$T_{c  n}= 6\times 10^8 \,\rm K$, 
and protons, $T_{c  p}=5\times 10^9 \,\rm K$. 
%
Let us stress that
the actual values of baryon critical temperatures 
as well as their density profiles are poorly known.
Different authors obtain different results for 
$T_{c n}$ and $T_{c p}$
depending on what microphysical input they use and how 
they take into account the many body effects \cite{yls99, ls01}.
Because of a large uncertainty in the models of baryon critical temperatures,
in this paper we will treat them as free parameters.
Our models of $T_{c n}$ and $T_{c p}$ do not contradict these,
available in the literature.

One sees from Fig.\ 1 that at low temperatures ($T \la 10^8$ K), 
the elements of the matrix $Y_{ik}$ 
are nearly independent of temperature 
and coincide with their values at $T=0$. 
With increasing temperature, 
they start to decrease 
and vanish after the corresponding critical temperature is reached. 
Thus, the matrix $Y_{ik}$ strongly depends 
on $T$ near the superfluid phase transition.

\begin{figure}[t]
\setlength{\unitlength}{1mm}
\leavevmode
\hskip  0mm
\includegraphics[width=100mm,bb=15 480 330 780,clip]{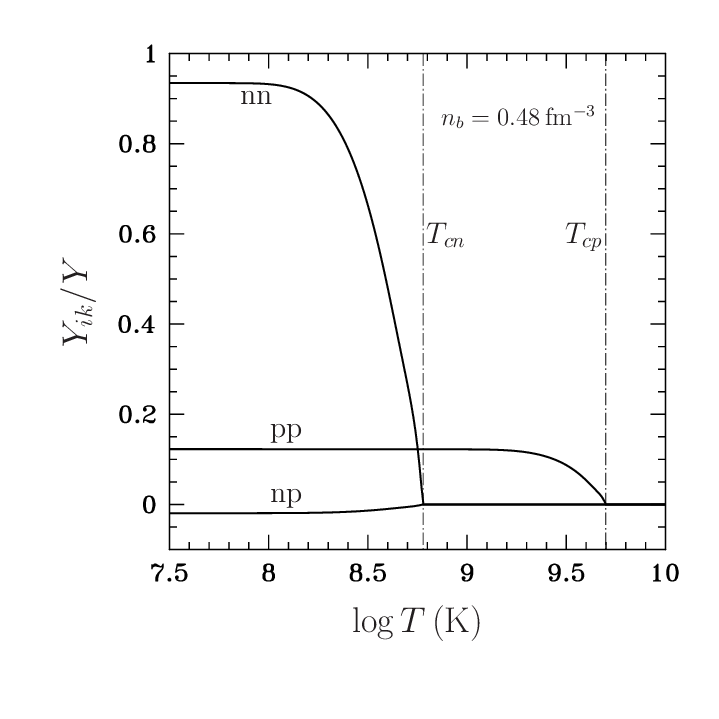}
\caption{
Normalized symmetric matrix $Y_{ik}/Y$ 
as a function of $T$ for $n_{ b}=3 n_0=0.48 \,\rm fm^{-3}$. 
The normalization constant $Y = 3n_0/\mu_{ n}(3n_0) 
\approx 2.60\times 10^{41}\, \rm erg^{-1}\,cm^{-3}$. 
Solid lines show the elements $Y_{ik}/Y$ ($i, k = n, p$). 
Vertical dot-dashed lines indicate the baryon critical temperatures 
$T_{c  n}=6\times 10^8 \,\rm K$ and $T_{c  p}=5\times 10^9 \,\rm K$.
}
\label{fig1}
\end{figure}
%

Along with the continuity equations and the energy-momentum conservation,
one should impose an additional constraint on the superfluid velocities,
following from the potentiality of superfluid motion \cite{ga06, gusakov07},
\begin{eqnarray}
\partial^{\nu} \left[ w^{\mu}_{(i)}
+q_i A^{\mu} + \mu_i u^{\mu} \right]
= \partial^{\mu} \left[ w^{\nu}_{(i)}
 +q_i A^{\nu} +\mu_i u^{\nu} \right],
\label{w_i2}
\end{eqnarray}
where $q_i$ is the charge of a species $i$, 
and $A^{\mu}$ is the four-potential of the electromagnetic field.

To close the system, the hydrodynamic equations should be supplemented 
with the second law of thermodynamics, 
which has the form 
\cite{ga06, gusakov07}
\begin{equation}
d \varepsilon =  T \, dS + \mu_e \, d n_e + \mu_i \, d n_i 
+ \frac{Y_{ik}}{2} \, d \left( w^{\alpha}_{(i)} w_{(k) \alpha} \right),
\label{2ndlaw}
\end{equation}
where $S$ is the entropy density.

\subsection{Radial pulsations}
The metric of a radially pulsating nonrotating 
symmetric NS has the form 
(see, e.g., \cite{chandrasekhar64}):
\begin{equation}
\dd s^2 =  -{\rm e}^{\nu} \dd t^2 + r^2 \dd \Omega^2
+ {\rm e}^{\lambda} \, \dd r^2,
\label{ds}
\end{equation}
where $r$ and $t$ are the radial and time coordinates, respectively; 
$\dd \Omega$ is the element of a solid angle 
in a spherical coordinate system with the origin at the stellar center. 
The metric coefficients $\nu(r, t)$ and $\lambda(r, t)$ 
depend only on $r$ and $t$. 
In what follows, the quantities related to a star 
in the hydrostatic equilibrium will be denoted by the subscript `0'. 
For example, the metric coefficients 
of an unperturbed star are designated as $\nu_0(r)$ and $\lambda_0(r)$.

Using Eqs.\ (1)--(5) the authors of GA06 derived 
the system of linear equations that describes radial pulsations 
of a general relativistic superfluid NS 
(all perturbations were assumed to be small 
and dependent on time $t$ as ${\rm exp}(i\omega t)$, 
where $\omega$ is the pulsation frequency):
\begin{eqnarray}
{\rm e}^{\lambda_0-\nu_0} \, \omega^2 \, n_{ b0} \mu_{ n0} \xi_{b} =
{\partial \delta P \over \partial r} + \delta P \, {\dd \over \dd r} 
\left( {1 \over 2} \lambda_0 + \nu_0 \right) + \nonumber\\
{1 \over 2} \, \delta \varepsilon 
\, 
{\dd \nu_0 \over \dd r} - {1 \over 2} \, n_{ b0} \mu_{ n0} \xi_{b} \, 
\left( {\dd \nu_0 \over \dd r} +{1 \over r} \right) \, {\dd \over \dd r} 
\left( \lambda_0 + \nu_0 \right),
\label{norm}\\
\mu_{ n0} \,\, {\rm e}^{\lambda_0-\nu_0/2} \,\omega^2 \,
\left( z_{ n} + \xi \right)
={\partial \over \partial r} 
\left( \delta \mu_{ n} \, {\rm e}^{\nu_0/2} 
+ {1 \over 2} \,\, \mu_{ n0} \, {\rm e}^{\nu_0/2}\, \delta \nu \right).
\label{sfl0}
\end{eqnarray}
Here, the quantities $\delta \varepsilon$, $\delta P$, 
$\delta \mu_n$, and $\delta \nu$ 
are, respectively, 
perturbations of 
$\varepsilon$, $P$, $\mu_n$ 
and the metric coefficient $\nu(r, t)$;
$\xi$ is the Lagrangian displacement 
of the normal liquid component, 
related to its velocity by \cite{chandrasekhar64}
\begin{equation}
u^1={\rm e}^{-\nu_0/2} \,\frac{\partial \xi}{\partial t}= 
i \, \omega \, {\rm e}^{-\nu_0/2} \, \xi.
\label{xi}
\end{equation}
Furthermore, $z_i$ is the Lagrangian displacement 
of a superfluid component $i$ relative to the normal component, defined as
\begin{equation}
w^1_{(i)}=\mu_{i0} \, {\rm e}^{-\nu_0/2} \,\frac{\partial z_i}{\partial t}
= i \, \omega \, \mu_{i0} \, {\rm e}^{-\nu_0/2} \, z_i,
\label{zi}
\end{equation}
and the function $\xi_{b}$ is 
proportional to the baryon current density 
$j^{\mu}_{(b)} = n_{ b} u^{\mu} + Y_{{ n}k} w^{\mu}_{(k)} 
+ Y_{{ p}k} w^{\mu}_{(k)}$ 
[see Eq.\ (\ref{currents})], and is given by
\begin{equation}
\xi_{ b} = \xi + \frac{\mu_{i0} Y_{{ n}i} \, z_i}{n_{ b0}}
+\frac{\mu_{i0} Y_{{ p}i} \, z_i}{n_{ b0}}.
\label{xib}
\end{equation}

Equation (\ref{norm}) is
the relativistic Euler equation generalized 
to the case of superfluid stars. 
A similar equation has been derived 
for normal matter by Chandrasekhar \cite{chandrasekhar64}. 
Equation (\ref{sfl0}), specific to superfluid matter, 
is a consequence of the potentiality of superfluid motion 
[see condition (\ref{w_i2})].

Because we are interested in oscillations 
whose frequencies are substantially smaller 
than the electron plasma frequency, 
one can assume that the quasineutrality condition 
$n_{ p}=n_{ e}$ is always satisfied. 
As a consequence, the electron and proton currents 
must be equal, $j^{\mu}_{( p)}=j^{\mu}_{( e)}$. 
Using now expression (\ref{currents}) 
and the definition (\ref{zi}), 
one obtains an additional constraint that relates $z_n$ and $z_p$,
\begin{equation}
\mu_{k0} \, Y_{{ p}k} \, z_{k} = 0.
\label{quasineutrality}
\end{equation}

In this study, it will be convenient 
to transform Eq.\ (\ref{sfl0}) 
and rewrite it in a somewhat different form. 
For this purpose, we multiply Eq.\ (\ref{sfl0}) 
by $n_{ b0}\,{\rm exp}(-\nu_0/2)$ 
and subtract Eq.\ (\ref{norm}) from the result. 
After some algebra one gets, using equation (54) 
for $\partial \delta \nu/\partial r$ from GA06
(see also \cite{gk10a,gk10b}),
\begin{eqnarray}
\mu_{ n0} \,\, n_{ b0} \,\, {\rm e}^{\lambda_0-\nu_0} \, \omega^2 \,
\left( z_{ n} -\frac{\mu_{k0} Y_{{ n}k} z_k}{n_{ b0}} \right)
=n_{ e0}\,\, {\partial \delta \mu \over \partial r},
\label{sfl}
\end{eqnarray}
where $\delta \mu \equiv \mu_{ n}-\mu_{ p}-\mu_{ e}$ 
is the disbalance of chemical potentials. 
Solving the system of equations (\ref{norm}), (\ref{quasineutrality}), and (\ref{sfl}), 
we can find the eigenfunctions $\xi$, $z_n$, and $z_p$, 
as well as the spectrum of eigenfrequencies $\omega$. 
In order to do so, it is, however, 
necessary to express the quantities 
$\delta \varepsilon$, $\delta P$, and $\delta \mu$ 
in terms of the functions $\xi$, $z_n$, and $z_p$ 
and to formulate the boundary conditions 
for Eqs. (\ref{norm}) and (\ref{sfl}).

Owing to the quasineutrality condition, $n_e=n_p$, 
any thermodynamic quantity can be presented 
as a function of two variables, say, $n_{b}$ and $n_{e}$ 
(the dependence 
on the quadratically small scalars 
$w^{\mu}_{(i)}w_{(k) \mu}$ and on $T$ can be neglected). 
Because we consider small-amplitude pulsations, 
the pressure perturbation and 
the chemical potential disbalance 
can be expanded in Taylor series 
near their equilibrium values,
\begin{eqnarray}
\delta P &=& {\partial P(n_{ b0}, n_{ e0}) \over \partial n_{ b0}} 
\, 
\delta n_{ b}
+{\partial P(n_{ b0}, n_{ e0}) \over \partial n_{ e0}} \,
\delta n_{ e},
\label{ExpandP} \\
\delta \mu &=& {\partial \delta \mu(n_{ b0}, n_{ e0}) \over 
\partial n_{ b0}} \, 
\delta n_{ b}
+{\partial \delta \mu (n_{ b0}, n_{ e0}) \over \partial n_{ e0}} 
\,
\delta n_{ e}.
\label{expanddmu}
\end{eqnarray}
Similarly, in view of the beta-equilibrium condition 
$\delta \mu_0=\mu_{n0}-\mu_{p0}-\mu_{e0}=0$, valid for an unperturbed star, 
one has from Eq.\ (\ref{2ndlaw}) (see also GA06),
\begin{equation}
\delta \varepsilon = \mu_{n0} \, \delta n_b.
\label{depsilon}
\end{equation}
Using continuity equations (\ref{currents}), 
the variations $\delta n_b$ and $\delta n_e$ 
appearing in Eqs.\ (\ref{ExpandP})--(\ref{depsilon}) 
can be expressed in terms of the Lagrangian displacements 
$\xi$, $z_n$, and $z_p$,
\begin{eqnarray}
\delta n_{ b} &=& -{{\rm e}^{\nu_0/2} \over r^2} \, {\partial \over \partial r} 
\left( r^2 \, n_{ b0} \, \xi_{ b}  \, {\rm e}^{-\nu_0/2} \right),
\label{dnb} \\
\delta n_{ e} &=& - {{\rm e}^{\nu_0/2} \over r^2} \, {\partial \over \partial r}
\left(  r^2 \,n_{ e0} \, \xi \, {\rm e}^{-\nu_0/2} \right).
\label{dne}
\end{eqnarray}
%
\subsection{Boundary conditions}

Prior to formulating the boundary conditions 
for Eqs. (\ref{norm}) and (\ref{sfl}), 
we should refine the model of a star whose pulsations we analyze. 
We consider a two-layer model constituted by two regions: 
an inner region, in which neutrons are superfluid, 
and an outer region, in which neutrons are normal.
In the inner region 
(for which $r \leq R_0$), 
the functions $z_i$ are nonzero; 
we name this region `superfluid'.  
In the outer region (for which $r>R_0$), 
all matter components move with the same velocity $u^{\mu}$, 
i.e., the fluid demonstrates 
a nonsuperfluid behavior 
[even if protons remain superfluid, 
it follows from (\ref{quasineutrality}) that $z_{ p}=0$]; 
we name such a region `normal'.  
In addition, we assume that the thermodynamic functions 
(e.g., the particle number densities $n_{l}$) 
do not experience any abrupt change 
at the superfluid-normal interface. 
Then one can formulate the following boundary conditions 
for such a stellar model (see also GA06).

(1) Vanishing of the pressure at the stellar surface.
\begin{equation}
\left[ \delta P + {\dd P_0 \over \dd r} 
\,  \xi_{ b}\right]_{r=R} = 0,
\label{boundary1} 
\end{equation}
where $R$ is the circumferential radius of the star.

(2) Existence of all the derivatives 
with respect to $r$ at the stellar center, 
which means that the following limits must be finite
\begin{equation}
\lim\limits_{r \rightarrow 0} \xi_{ b}/r < \infty, \quad \quad
\lim\limits_{r \rightarrow 0} \mu_{i0} Y_{{ n}i} \, z_i/r< \infty.
\label{boundary23} 
\end{equation}

(3) Continuity of the energy current density across 
the superfluid-normal interface.
As shown in GA06, this current is directed along the radial coordinate
and is proportional to 
$n_{ b0} \mu_{ n0} \xi 
+ \mu_{ n0} \mu_{i0} Y_{{ n} i} z_i = n_{ b0} \mu_{ n0} \xi_{ b}$. 
Hence, because of the continuity 
of the thermodynamic functions, one has
\begin{equation}
\left[ \xi_{ b} \right]_{r=R_0-0}
= \left[ \xi_{ b} \right]_{r=R_0+0}. 
\label{boundary4} 
\end{equation}

(4) Continuity of the electron current density 
across the superfluid-normal interface, 
$\xi(R_0-0) =\xi(R_0+0)$, 
which, combined with Eqs.\ (\ref{xib}), (\ref{quasineutrality}), 
and (\ref{boundary4}), leads to
\begin{equation}
\mu_{i0} Y_{{ n}i} \, z_i \mid_{r=R_0} = 0.
\label{boundary5}  
\end{equation}
The solution to the system of equations (\ref{norm}), 
(\ref{quasineutrality}), and (\ref{sfl}) 
must satisfy these four boundary conditions.

\section{Pulsation spectra}
\label{3}

Using Eqs.\ (\ref{norm}), (\ref{quasineutrality}), and (\ref{sfl}), 
we determined the spectrum of eigenfrequencies $\omega$ 
of a radially pulsating NS as a function of its internal temperature $T$. 
In our numerical calculations, 
we considered NS with the mass $M=1.4 M_{\odot}$, 
circumferential radius $R = 12.17$ km, 
and central density $\rho_{\rm c} = 9.26 \times 10^{14}$ g cm$^{-3}$. 
We used 
APR EOS in the core, 
and Negele and Vautherin EOS in the crust \cite{nv73}. 
For the chosen model of a star, the core-crust interface 
lies at a distance $R_{\rm cc} = 10.88$ km from the stellar center. 
Following GA06, free neutrons in the inner crust were assumed 
to be normal, while in the core 
both nucleon species were allowed to be superfluid.

To simplify calculations, 
the red-shifted critical temperature of protons 
was chosen to be constant throughout the core 
(i.e., independent of the density $\rho$) 
and equal to 
$T_{c  p}^\infty \equiv T_{c  p}\rm e^{\nu_0/2}=5\times 10^9$ K. 
Note that, the pulsation spectrum 
only weakly depends on a particular form of the function 
$T_{c  p}^\infty(\rho)$ and even on whether or not protons are superfluid. 
The point is the only quantity in superfluid hydrodynamics,
which depends on the critical temperatures $T_{cn}$ and $T_{cp}$,
is the entrainment matrix $Y_{ik}$.
%
%
The element $Y_{nn}$ of the matrix is mostly determined 
by the neutron critical temperature $T_{cn}$ 
and is nearly insensitive to $T_{cp}$ \cite{gkh09b}. 
By contrast, the elements $Y_{pp}$ and $Y_{np}$ 
strongly 
vary with $T_{cp}$,
but the terms, 
depending on these matrix elements,
are small in pulsation equations 
and can be neglected to a good approximation.

To illustrate this statement let us notice that, 
as follows from Eqs. (\ref{norm}) and (\ref{sfl}), 
$Y_{np}$ appears there only
in the combination $\mu_n \, Y_{nn} \, z_n + \mu_p \, Y_{np} \, z_p$ 
(and $Y_{pp}$ does not explicitly enter these equations). 
From the quasineutrality condition (\ref{quasineutrality}), 
we obtain $z_p= -\mu_n \, Y_{np} \, z_n/(\mu_p Y_{pp})$, 
and, therefore, the second term in this combination differs 
from the first one by a factor $Y_{np}^2/(Y_{nn}Y_{pp})$. 
As follows from Fig.\ 1, 
$Y_{np}^2/(Y_{nn}Y_{pp}) \sim 0.01$ 
and, consequently, this term can be indeed neglected.

The red-shifted temperature of unperturbed star, 
$T^\infty=T\rm e^{\nu_0/2}$, 
was assumed to be finite and constant throughout the stellar core. 
The latter is a necessary condition 
for a superfluid star to be in the hydrostatic equilibrium 
[see equation (30) of GA06]. 
The influence of finite internal temperature 
on the pulsations of NSs is twofold.
First, 
it determines 
the entrainment matrix $Y_{ik}(T)$ 
(see Fig.\ 1). 
Second, 
it regulates the size $R_0$
of the 
superfluid 
region, 
for which $T \leq T_{c n}(\rho)$:
the higher the temperature, 
the smaller the region 
(see Fig.\ 2).

We considered four models 
showing different behavior 
of $T_{c  n}(\rho)$
and allowing for 
one or both of these temperature effects.

Model 1. The red-shifted neutron critical temperature $T_{c  n}^\infty$ 
is density independent and equals
$T_{c  n}^\infty=T_{c  n}\rm e^{\nu_0/2}=6\times 10^8$~K. 
This model assumes that
superfluid region coincides 
with the stellar core at $T^\infty \leq T_{cn}^\infty$.
Within this region we use the finite-temperature 
hydrodynamics, i.e., take into account 
the full dependence of $Y_{ik}$ on $T$.
This model is identical to the model considered in GA06. 
The only difference is that the authors of GA06 
used an approximate expression for the matrix $Y_{ik}$ 
[see their equation (17)], 
whereas in the present study 
we use results of a self-consistent 
relativistic calculation, performed in \cite{gkh09b}. 
However, one can check that 
the expression
for $Y_{ik}$ from GA06 
is a good approximation 
for calculation of the pulsation spectrum.

Model 2. 
$T_{c  n}^\infty$ depends on density and has a maximum at the stellar center, 
$T_{cn}^\infty(r=0)=6\times 10^8 \rm \,K$. 
The size of the superfluid region is temperature dependent 
and the boundary $R_0$ of this region is determined 
by the condition 
%
\begin{equation}
T_{c  n}^\infty(R_0)=T^\infty. 
\label{R0}
\end{equation}
However, within the superfluid region 
we use the zero-temperature hydrodynamics, 
that is, the entrainment matrix 
in this model does not depend on $T$,
$Y_{ik}=Y_{ik}(T=0)$. 
This model has been used in a number of studies 
\cite{lac08,hap09,ha10}. 
In particular, in Ref.\ \cite{lac08} 
it was claimed that the model adequately 
accounts for the main temperature effects.

Model 3. The same as model 2, but within the superfluid region 
we use a hydrodynamics which is valid at finite temperatures; 
$Y_{ik}$ is temperature dependent.

Model 4. The same as model 3, 
but with 
$T_{c  n}^\infty$ having a maximum 
at a density exceeding the central density $\rho_c$. 
At the stellar center, $T_{c  n}^\infty(0)=6\times 10^8$~K.

It should be stressed that,
for each of these models, 
the results obtained
below in this paper
will remain {\it qualitatively} the same 
if we choose another neutron critical temperature $T^\infty_{cn}(0)$ 
in the stellar center
or if we
vary
the slope of the density profile $T_{cn}^\infty(\rho)$ 
(for models 2, 3, and 4).
The only important thing that 
discriminates between the models and that
should not be altered
is the condition 
$[{\rm d} T^\infty_{cn}(\rho)/{\rm d}\rho] |_{r=0} = 0$ 
for models 2 and 3 and 
$[{\rm d} T^\infty_{cn}(\rho)/{\rm d}\rho] |_{r=0} > 0$ 
for model 4. 
As it will be demonstrated in Secs.\ III and IV, 
this condition influences the pulsation spectra 
and the asymptotes for eigenfrequencies
at $T^\infty \rightarrow T_{cn}^{\infty}(0)$.

For illustration, 
Fig.\ 2 shows the red-shifted
$T_{c  n}^\infty$ (dashed curve) 
and the local $T_{c  n}$ (solid curve) neutron critical temperatures 
as function of density $\rho$
(left panel for models 2 and 3; right panel for model 4). 
Dot-dashed lines indicate the central density 
of the star and the density at the crust-core interface.
\begin{figure}[t]
\setlength{\unitlength}{1mm}
\leavevmode
\hskip  0mm
\includegraphics[width=150mm,bb=20 40 550 310,clip]{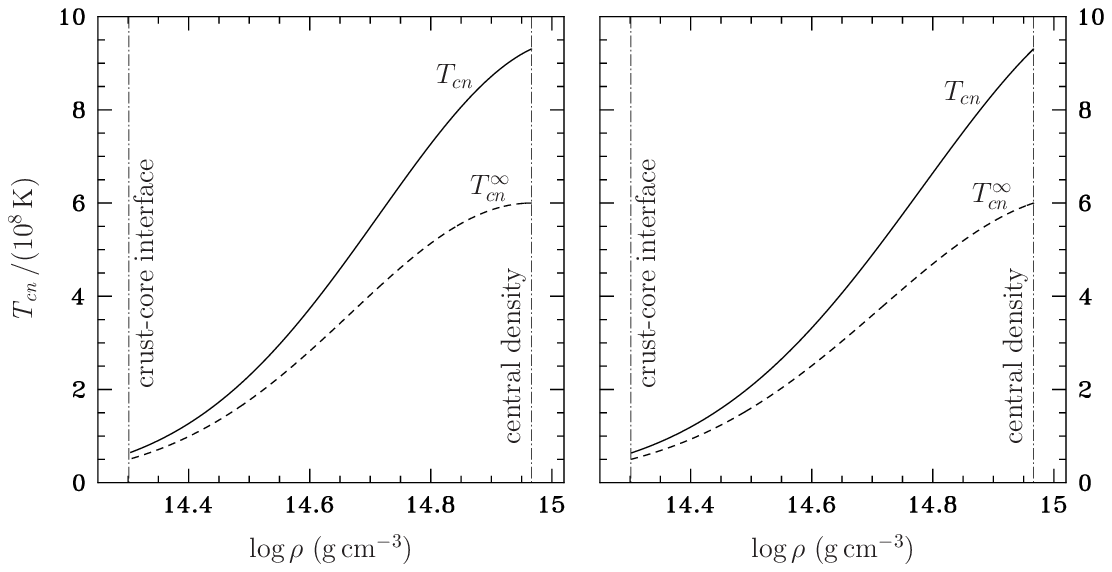}
\caption{Neutron critical temperatures $T_{c  n}^\infty$ (dashed lines) 
and $T_{c  n}$ (solid lines) versus $\rho$ 
(left panel for models 2 and 3, right panel for model 4). 
Dot-dashed lines show the central density and crust-core interface.
}
\label{fig11}
\end{figure}
%
Notice that, for all the models, a sufficiently cold star 
(at $T^\infty \leq 6\times 10^8\, \rm K$) 
consists of superfluid inner and normal outer regions. 

In Fig.\ 3, for each of the four models, 
we present an eigenfrequency spectrum 
of a radially pulsating superfluid NS. 
The alternate solid and dashed lines 
show eigenfrequencies $\omega$ 
(in units of $\omega_0 \equiv c/R$) as functions of $T^\infty$
for the first six pulsation modes. 
The dot-dashed line indicates the red-shifted neutron 
critical temperature at the stellar center, 
$T^{\infty}_{c  n}(0)=6\times10^8$ K. 
The dotted lines turning into thin solid lines 
at a temperature $T^{\infty} > T^{\infty}_{cn}(0)$ 
show first three radial oscillation modes 
of a normal star (I, II, and III).
The spectrum was not plotted in the shaded region.

It can be seen from the figure 
that at $T^{\infty} < T^{\infty}_{c  n}(0)$ 
there exist two types of oscillations.
Oscillation frequencies of the first type
are nearly temperature independent 
and well coincide with the corresponding eigenfrequencies 
of a nonsuperfluid star of the same mass 
(we name this type of modes `normal').
Oscillation frequencies of the second type 
strongly vary with temperature  
(for $T \ga 5 \times 10^7 - 10^8$ K); 
the corresponding modes are named `superfluid'.
With growing $T$, 
neighboring modes become closer and closer
until they form 
avoided crossing. 
Near such avoided crossings any superfluid mode 
turns into a normal mode and vice versa. 
At low enough temperatures 
($T \la 5\times10^7$ K for models 2, 3, and 4, 
and $T \la 10^8$ K for model 1), 
the eigenfrequencies cease to depend on $T$
and reach their asymptotic values.
The reason for that is, 
first, the matrix $Y_{ik}$ becomes temperature independent 
at $T^\infty \la 10^8$ K and second, 
the radius of the superfluid region reaches 
its maximum value (corresponding to the core-crust interface) 
at $T^\infty = 5\times10^7$ K and stops to vary 
as $T$ is lowered further.

For the superfluid type of pulsations, 
the decrease in the size $R_0$ of the superfluid region 
with increasing $T^\infty$ leads to a rise 
of the eigenfrequencies $\omega$ 
(this is clearly seen for model 2 
and stems from the fact that, 
as shown below, $\omega \propto 1/R_0$). 
At the same time, 
$Y_{nk} \rightarrow 0$ 
with increasing $T$ (Fig.\ 1), 
which results in decreasing of $\omega$ 
(this is exemplified by model 1 
and follows from the fact that 
$\omega \propto \sqrt{Y_{nn}}$ 
at $T^\infty \rightarrow T_{cn}^\infty(0)$, see below).
In the models 3 and 4, these two effects compete. 
Asymptotes for the eigenfrequencies at $T^{\infty}\rightarrow T^{\infty}_{c  n}(0)$ 
will be studied in detail in Sec.\ IVB. 
At temperatures higher than $T^{\infty}_{c  n}(0)$, 
the star pulsates as a normal one 
[even if protons still remain superfluid, 
all the liquid components move together by virtue 
of the quasineutrality condition (\ref{quasineutrality})].

As follows from Fig.\ 3, 
the pulsation spectra
differ substantially for different models.
For example, 
for most realistic models 3 and 4 
in the limit $T^{\infty}\rightarrow T^{\infty}_{c  n}(0)$,
the eigenfrequencies either vanish (model 3), 
or decrease, approaching some certain finite values (model 4).
The eigenfrequencies for model 1 
behave similar to those for model 3. 
At the same time, 
the eigenfrequencies for model 2 behave in a drastically different way. 
At $T^{\infty}\rightarrow T^{\infty}_{c  n}(0)$
they grow rather than decrease
and 
tend to infinity as $R_0 \rightarrow 0$. 
This example clearly demonstrates that it is insufficient 
just to vary the size of the superfluid region with $T$,
using the zero-temperature hydrodynamics (as in model 2). 
Such an approach gives {\it qualitatively incorrect} spectra. 
Thus, it is of principal importance
to take into account 
the temperature dependence of the entrainment matrix.

\begin{figure}[t]
\setlength{\unitlength}{1mm}
\leavevmode
\hskip  0mm
\includegraphics[width=160mm,bb=15 80 495 530,clip]{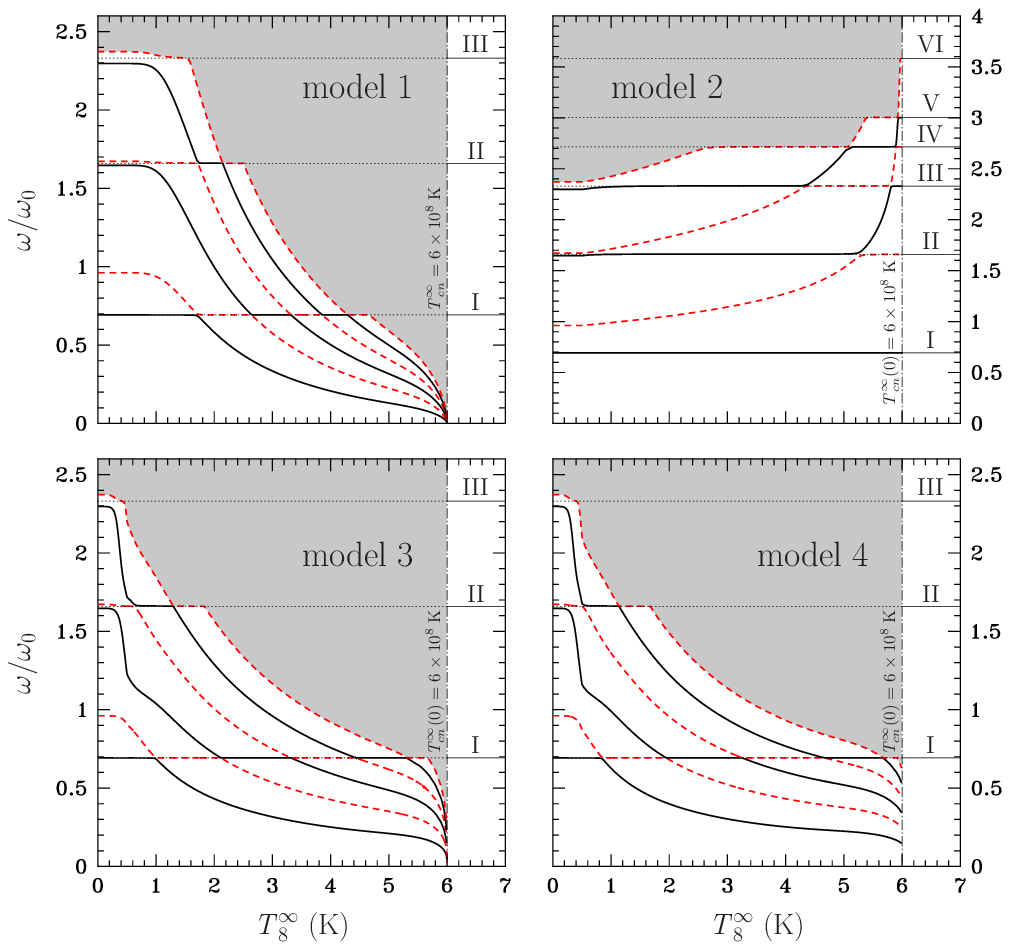}
\caption{
Eigenfrequencies $\omega$ (in units of $\omega_0 = c/R$) 
versus $T^{\infty}_8=T^{\infty}/10^8$, 
calculated for models $1$, $2$, $3$, and $4$ (see the text). 
Solid and dashed curves show 
the first six pulsation modes. 
The red-shifted neutron critical 
temperature $T^{\infty}_{c  n}(0)$ 
in the stellar center
is indicated by the vertical 
dot-dashed lines; 
the horizontal dotted
lines turning into solid lines at 
$T^{\infty} > T^{\infty}_{c  n}(0)$
show the first three eigenfrequencies 
(I, II, III) 
of a normal star of the same mass 
(notice that, for model 2 
we show the first {\it six} eigenfrequencies of a normal star). 
No spectrum was plotted in the shaded region.}
\label{fig111}
\end{figure}

\section{Eigenfrequencies at $T^\infty \rightarrow T_{cn}^\infty(0)$}

\subsection{Approximate splitting of pulsation equations}

As shown in Refs.\ \cite{gk10a,gk10b}, 
the equations that govern global pulsations of superfluid NSs 
can generally be represented 
as two weakly coupled systems of equations. 
The coupling parameter $s$ for these systems is given by
\begin{equation}
s=\frac{n_{ e0}}{n_{ b0}}\,\,
\frac{\partial P(n_{ b0}, n_{ e0}) 
/ \partial n_{ e0}}{\partial P(n_{ b0}, n_{ e0}) 
/ \partial n_{ b0}}
\label{s}
\end{equation}
%
\begin{figure}[t]
\setlength{\unitlength}{1mm}
\leavevmode
\hskip  0mm
\includegraphics[width=80mm,bb=10 450 330 750,clip]{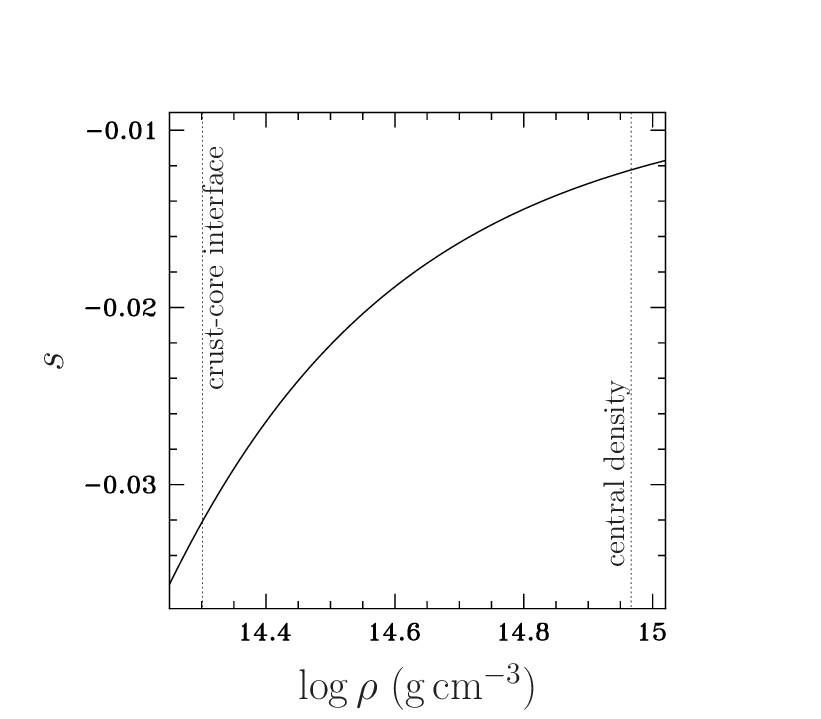}
\caption{Coupling parameter $s$ 
versus $\rho$ for APR EOS.
The vertical dotted lines indicate 
the central stellar density 
and the crust-core interface.}
\label{sfig}
\end{figure}
and is small for realistic equations of state, 
$|s| \sim 0.01-0.05$. 
In particular, $s$ for APR EOS is shown in Fig.\ 4. 
In the case of radial pulsations, 
the smallness of the parameter $s$ 
means that the pulsation equations (\ref{norm}) and (\ref{sfl}) 
can be treated as independent 
to a very good approximation
(see \cite{gk10a,gk10b}). 
If we assume $s = 0$, then Eq. (\ref{norm}) 
can be separated from Eq.\ (\ref{sfl}) 
and written only in terms of the variable $\xi_{ b}$:
\begin{eqnarray}
{\rm e}^{\lambda_0-\nu_0} \, \omega^2 \, n_{ b0} \mu_{ n0} \xi_{ b} =
{\partial \delta P_{\rm norm} \over \partial r} + \delta P_{\rm norm} \, {\dd \over \dd r} 
\left( {1 \over 2} \lambda_0 + \nu_0 \right) \nonumber\\
+ {1 \over 2} \, \delta \varepsilon 
\, 
{\dd \nu_0 \over \dd r} - {1 \over 2} \, n_{ b0} \mu_{ n0} \xi_{ b} \, 
\left( {\dd \nu_0 \over \dd r} +{1 \over r} \right) \, {\dd \over \dd r} 
\left( \lambda_0 + \nu_0 \right),
\label{normapp}
\end{eqnarray}
where
\begin{eqnarray}
\delta P_{\rm norm} \equiv {\partial P(n_{ b0}, n_{ e0}) \over \partial n_{ b0}} 
\, \delta n_{ b} + {\partial P(n_{ b0}, n_{ e0}) \over \partial n_{ e0}} 
\, \delta n_{ e\,\rm norm}, \nonumber\\
\delta n_{ e\,\rm norm} \equiv - {{\rm e}^{\nu_0/2} \over r^2} \, {\partial \over \partial r}
\left(  r^2 \,n_{ e0} \, \xi_{ b} \, {\rm e}^{-\nu_0/2} \right).
\label{dPnorm}
\end{eqnarray}
In this form, the equation fully coincides 
with that describing oscillations of a nonsuperfluid star. 
Its solutions are eigenfrequencies $\omega$ 
and eigenfunctions $\xi_{ b}$ of a normal star. 
Thus, Eq.\ (\ref{normapp}) approximately describes 
the normal modes of oscillations in our problem. 
As for the superfluid modes, 
they are described by an equation 
depending only on `superfluid' Lagrangian 
displacements $z_n$ and $z_p$ \cite{gk10a,gk10b},
\begin{equation}
\mu_{ n0} n_{ b0}\, {\rm e}^{\lambda_0-\nu_0}
\omega^2 \, \left( z_{ n}- \frac{\mu_{k0} Y_{nk} \, z_{ k}}{n_{ b0}} \right)
=n_{ e0} \,\, \frac{\partial \delta \mu_{\rm sfl}}{\partial r},
\label{sflapp}
\end{equation}
where
\begin{equation}
\delta \mu_{\rm sfl}\equiv \frac{\partial \delta \mu}
{\partial n_{ e}}\, \frac{{\rm e}^{\nu_0/2}}{r^2}\,
{\partial \over \partial r}\left(  r^2 \,\frac{n_{ e0}}{n_{ b0}} 
\, \mu_{i0} Y_{{ n}i} z_i \, {\rm e}^{-\nu_0/2} \right).
\label{musfl}
\end{equation}
Eq.\ (\ref{sflapp}) can be derived from (\ref{sfl}) 
if we put $\xi_b=0$ in the latter. 
Indeed, if some frequency is an eigenfrequency 
for the system of equations (\ref{sfl}) and (\ref{normapp}), 
but is not an eigenfrequency 
for Eq.\ (\ref{normapp}), 
then this is only possible if $\xi_b$, 
the eigenfunction of Eq.\ (\ref{normapp}), 
identically vanishes. 
(This consideration is strictly valid when $s=0$. 
For small but finite $s$ one has
$\xi_{ b} \ll \xi \approx -\mu_{i0} Y_{{ n}i} \, z_i/n_{ b0}$.)

The approximation of independent Eqs.\ (\ref{normapp}) and (\ref{sflapp}) 
well reproduces (with an accuracy of $1.5-2 \%$) 
the frequency spectrum of a pulsating NS. 
However, this approximation does not describe 
avoided crossings of neighboring modes 
(see Fig.\ 3).
The reason is near these points 
Eqs. (\ref{norm}) and (\ref{sfl}) 
are strongly interacting \cite{gk10a, gk10b}, 
so that 
approximate treatment based on 
the decoupled Eqs. (\ref{normapp}) and (\ref{sflapp}) 
is unjustified;
instead of avoided crossings, 
Eqs. (\ref{normapp}) and (\ref{sflapp}) predict crossings of modes. 
It should be noted that the existence of two types of modes, 
superfluid and normal, 
has been noted previously by a number of authors 
(see, e.g., Refs.\ \cite{lm94,pr02, hap09}).
However, an adequate analytical interpretation 
this fact first received in Ref.\ \cite{gk10a}.

\subsection{Asymptotes for eigenfrequencies}

Using the $s = 0$ approximation described above, 
let us consider the asymptotic behavior 
of eigenfrequencies at $T^\infty \rightarrow T_{c  n}^\infty (0)$. 
For all of the four models, 
any given mode experiences superfluid-type pulsations
near the neutron critical temperature
and, therefore, to derive the asymptotes 
it is necessary to use Eq.\ (\ref{sflapp}). 

As the temperature approaches the critical value $T_{c  n}$, 
the elements $Y_{{ n}i}(\rho, T/T_{cn}, T/T_{cp})$ 
of the entrainment matrix tend to vanish according to the law 
\cite{gkh09b, gh05}: 
$Y_{{ n}i}=F_{ni} \, (1-T/T_{c n})$, 
where $F_{ni}$ is some (known) function of 
$\rho$, $T/T_{cn}$, and $T/T_{cp}$, 
which are in turn the functions of radial coordinate $r$.
Taking into account that $T/T_{ci}=T^\infty/T^{\infty}_{ci}$, 
the asymptote for $Y_{ni}$ can be rewritten as
\begin{equation}
Y_{{ n}i}=F_{ni}(r,T^\infty) (1-T^\infty/T_{c n}^\infty) \rightarrow 0.
\label{Yikasy}
\end{equation}
Owing to the smallness of the matrix elements $Y_{{ n}i}$
near the neutron critical temperature, 
one can {\it neglect} the second term
in the left-hand side of Eq.\ (\ref{sflapp}) 
when analyzing asymptotes for models 1, 3, and 4.

In addition, 
the size $R_0$ of the superfluid region in models 2, 3, and 4
decreases with increasing temperature 
and vanishes at 
$T^\infty = T_{c  n}^\infty (0)$. 
The limiting behavior $R_0(T^\infty)$
can be found from the condition (\ref{R0})
and is determined by the profile 
of the neutron critical temperature. 
Expanding $T_{cn}^\infty(\rho)$ in the Taylor series 
in the vicinity of the stellar center 
and taking into account that 
for models 2 and 3 
$[{\rm d}T_{cn}^\infty/{\rm d}\rho]|_{r=0}=0$  
while for model 
4 $[{\rm d}T_{cn}^\infty/{\rm d}\rho]|_{r=0} \neq 0$, 
one gets, for models 2 and 3,
\begin{equation}
T_{c  n}^\infty(r)-T_{c  n}^\infty(0) 
\propto [\rho(r)-\rho_c]^2\propto r^4,
\label{Tcnprop23}
\end{equation}
and for model 4,
\begin{equation}
T_{c  n}^\infty(r)-T_{c  n}^\infty(0) 
\propto \rho(r)-\rho_c\propto r^2.
\label{Tcnprop4}
\end{equation}
To obtain the asymptotes (\ref{Tcnprop23}) and (\ref{Tcnprop4}) 
we used the fact that in the stellar center 
$[{\rm d}\rho/{\rm d} r]|_{r=0} = 0$. 
Now, using Eqs.\ (\ref{Tcnprop23}) and (\ref{Tcnprop4}) 
and the condition (\ref{R0}),
one finds how $R_0$ scales with $T^\infty$ 
at $T^\infty \rightarrow T_{cn}^\infty(0)$. 
For models 2 and 3 
\begin{equation}
R_0 \propto \left[1- T^{\infty}/T_{cn}^\infty(0) \right]^{1/4},
\label{R0prop23}
\end{equation} 
while for model 4
\begin{equation}
R_0 \propto \left[1- T^{\infty}/T_{cn}^\infty(0)\right]^{1/2}.
\label{R0prop4}
\end{equation}

Making use of Eqs.\ (\ref{Yikasy})--(\ref{R0prop4}), 
it is possible to analyze the dependence $\omega(T^\infty)$
at $T^\infty \rightarrow T_{cn}^{\infty}(0)$
for different models.
We start with the model 1.
Then, the only quantities in Eq.\ (\ref{sflapp}) 
that depend on $T$
are the matrix elements $Y_{{ n} i}$ 
and the frequency $\omega$
(we remind that 
the size $R_0$ of the superfluid region 
is fixed for model 1).
Using Eq.\ (\ref{Yikasy}) and the fact that 
$T^\infty/T^\infty_{cn}$ does not depend on $r$ for this model,
one can factor the temperature dependence 
of $Y_{{ n} i}$ outside the differentiation sign 
on the right-hand side of Eq.\ (\ref{sflapp}). 
The result is
\begin{equation}
\mu_{ n0} n_{ b0}\, {\rm e}^{\lambda_0-\nu_0}
\omega^2 \,  z_{ n} 
=n_{ e0} \left(1-\frac{T^\infty}{T_{c n}^\infty}\right)\,\, 
\frac{\partial }{ \partial r} \left[\frac{\partial \delta \mu}
{\partial n_{ e}}\, \frac{{\rm e}^{\nu_0/2}}{r^2}\,
{\partial \over \partial r}\left(  r^2 \,\frac{n_{ e0}}{n_{ b0}} 
\, \mu_{i0} F_{ni}(r,T^\infty) z_i \, {\rm e}^{-\nu_0/2} \right)\right].
\label{sflappfac}
\end{equation}
The function $F_{ni}(r,T^\infty) \approx F_{ni}(r, T_{cn}^\infty)$ 
in the immediate vicinity of $T_{cn}^\infty$. 
Taking this into account, one immediately obtains 
from Eq.\ (\ref{sflappfac}),
\begin{equation}
\omega \propto \sqrt{1-T^\infty/T_{c n}^\infty} \rightarrow 0, \quad 
T^\infty \rightarrow T_{cn}^\infty.
\label{w1}
\end{equation}

Now let us inspect model 2.
This model assumes that the matrix $Y_{ik}$ 
is temperature independent, 
while $R_0$ decreases with increasing $T$. 
In that case, the asymptote for the function $\omega(T^\infty)$ 
can be determined by introducing a new dimensionless variable, 
$\tilde{r}\equiv r/R_0$, 
which varies within the range from $0$ to $1$. 
Rewriting Eq.\ (\ref{sflapp}) 
employing the new variable, one finds 
[notice that we retain the second term 
on the left-hand side of Eq.\ (\ref{sflapp}) 
for this model]
\begin{equation}
\mu_{ n0} n_{ b0} \, {\rm e}^{\lambda_0-\nu_0} \omega^2\,R_0^2 
\left( z_{ n}- \frac{ \mu_{k0} Y_{nk} \, z_{ k}}{n_{ b0}} \right)
=n_{ e0} \, \frac{\partial}{\partial \tilde{r}}\,\left[ \frac{\partial \delta \mu}{\partial n_{ e}}\, \frac{{\rm e}^{\nu_0/2}}{\tilde{r}^2}\,{\partial \over \partial \tilde{r}}\left(  \tilde{r}^2 \,\frac{n_{ e0}}{n_{ b0}} \, \mu_{i0} Y_{{ n}i} z_i \, {\rm e}^{-\nu_0/2} \right) \right].
\label{model4}
\end{equation}
At $T^\infty$ sufficiently close to $T^\infty_{cn}(0)$, 
when the size $R_0$ of the superfluid region 
is rather small, 
such functions 
as $\mu_{i0}$, $Y_{ik}$, $n_{i0}$, and $\nu_0$ 
are almost independent of the radius $r$ 
[so that one can put, e.g., $\mu_{i0}(r) \approx \mu_{i0}(0)$].
Thus, the only quantities varying with temperature 
in Eq.\ (\ref{model4}) are $\omega$ and $R_0$, 
and, therefore, the asymptote takes the form
\begin{equation}
\omega \propto \frac{1}{R_0} \propto \frac{1}
{\left[1- T^{\infty}/T_{cn}^\infty(0) \right]^{1/4}} \rightarrow \infty, 
\quad T^\infty \rightarrow T_{cn}^\infty(0).
\label{model2}
\end{equation}
To obtain this asymptote we also used Eq.\ (\ref{R0prop23}). 

Finally, let us consider models 3 and 4.
In these models, both $Y_{ik}$ and $R_0$ 
depend on $T^\infty$. 
So, we need to combine our reasoning 
for the first and second models. 
The pulsation equation for models 3 and 4
is given by the same Eq.\ (\ref{model4}) 
but without the second term $\propto Y_{ni}$
on its left-hand side; 
as it is argued above in this section, 
this term is small at $T^\infty \rightarrow T_{cn}^\infty$
and can be omitted.
Again, as for model 2, 
in Eq.\ (\ref{model4}) one can replace
such quantities as $\mu_{i0}(r)$ 
by their values $\mu_{i0}(0)$ in the stellar center.
The same, of course, cannot be done for the matrix element 
$Y_{ni}(r, T^\infty)= F_{ni}(r, T^\infty) \, [1-T^\infty/T_{cn}^\infty(r)]$,
because it changes substantially in the superfluid region.
Using Eqs.\ (\ref{R0}) and (\ref{Yikasy})--(\ref{Tcnprop4}) it is easy to verify
that $Y_{ni}(r, T^\infty)$ can generally be presented as
\begin{equation}
Y_{ni}=F_{ni}(0, T_{cn}^\infty(0)) \,\, [1-T^\infty/T_{cn}^\infty(0)] \,\, (1-\tilde{r}^4) 
+ O([1-T^\infty/T_{cn}^\infty(0)]^2)
\label{Yniasym3}
\end{equation}
for model 3 and as
\begin{equation}
Y_{ni}=F_{ni}(0, T_{cn}^\infty(0)) \,\, [1-T^\infty/T_{cn}^\infty(0)] \,\, (1-\tilde{r}^2) 
+ O([1-T^\infty/T_{cn}^\infty(0)]^2)
\label{Yniasym4}
\end{equation}
for model 4.
The temperature dependence of $Y_{ni}$ 
can now be easily factored out 
of the differentiation sign 
on the right-hand side of Eq.\ (\ref{model4}).
In this way, one obtains the following asymptote for $\omega$,
\begin{equation}
\omega \propto \frac{\sqrt{1-T^\infty/T_{cn}^\infty(0)}}{R_0}.
\label{realapp}
\end{equation}
Now, employing Eqs.\ (\ref{R0prop23}) and (\ref{R0prop4}),
one finds for model 3
\begin{equation}
\omega \propto \left(1- \frac{T^{\infty}}{T_{cn}^\infty(0)}\right)^{1/4} \rightarrow 0,
\quad T^\infty \rightarrow T_{cn}^\infty(0),
\label{model33}
\end{equation}
and for model 4
\begin{equation}
\omega \rightarrow {\rm finite}, \quad T^\infty \rightarrow T_{cn}^\infty(0).
\label{model44}
\end{equation}
It should be emphasized that, 
owing to a weak dependence 
of the function $F_{ni}(r,T^\infty)$ on $T^\infty$, 
the eigenfrequencies for the model 4 
do not tend to a constant, 
but remain weakly dependent on temperature.
One can verify that the asymptotes obtained for all the models 
well describe our numerical results presented in Fig.\ 3. 

In this study, we only consider profiles $T_{c  n}^\infty(\rho)$ 
that have a maximum at densities higher than, or equal to
the central stellar density $\rho_c$. 
This enables us to consider only two regions in the star: 
inner superfluid and outer normal. 
However, a situation in which the maximum lies 
at $\rho < \rho_c$ can easily be imagined. 
In this case, there will be generally three regions in a star:
inner and outer normal regions and a superfluid region in between.
The size $R_0$ of the superfluid region
will decrease with increasing temperature 
in the same manner as in model 4, 
$R_0 \propto \left(1- T^{\infty}/T_{cn \, 
{\rm max}}^\infty\right)^{1/2}$ 
(where $T_{cn \,{\rm max}}^\infty$ 
is the value of $T_{cn}^\infty$ at the maximum). 
Therefore, at $T^\infty \rightarrow T_{cn \, {\rm max}}^\infty$
the eigenfrequencies for such star will have 
the same asymptotes as for the model 4, 
$\omega \rightarrow {\rm finite}$. 
The spectrum for a three-layer star 
is analyzed in Ref.\ \cite{cg11}.

\section{Eigenfunctions}
\label{4}

In this section we compare the main properties
of superfluid and normal pulsation modes.
For that we analyze the eigenfunctions 
for both types of modes. 
For brevity, 
we consider only the third model.

Figure 5 shows the eigenfunctions 
$\xi$, $z_{ n}$, $z_{ p}$, and $\xi_{ b}$ 
(normalized to the stellar radius $R$), 
calculated at $T^{\infty}=1.2\times10^8 \,\rm K$ 
versus $r$
(in units of $R$). 
The oscillation amplitude was chosen 
in such a way 
that the energy of radial pulsations 
is $E_{\rm puls} =10^{48}$ erg 
[see Eq.\ (\ref{pulsation_energy}) for the definition of $E_{\rm puls}$]. 
For a fundamental mode of a normal star this energy 
corresponds to an 
oscillation amplitude 
$\xi \sim 10^{-3} r$.
It could be also interesting to note 
that this energy is of the order of the thermal energy 
of a normal star with $T^\infty=10^9$~K \cite{gyg05}.
Each mode (I,...,VI) in Fig.\ 5 is plotted in a separate panel.
The boundary $R_0$ between the superfluid 
and normal regions,
and the crust-core boundary $R_{\rm cc}$,
are indicated by vertical dotted lines. 
The solid line corresponds 
to the Lagrangian displacement $\xi$ 
of the nonsuperfluid component; 
the dashed and dot-dashed lines 
correspond to relative Lagrangian displacements 
$z_{ n}$ and $z_{ p}$, respectively; 
the heavy solid line -- to the Lagrangian displacement $\xi_{ b}$, 
which is proportional to the baryon current density. 
We do not plot $z_{ n}$ 
in the nonsuperfluid region of the star
since all neutrons are unpaired there.
We also do not plot $z_{ p}$ in the crust,
because it contains no free protons. 
The quantity $\xi_{ b}$ in the normal region 
of the star coincides with $\xi$.

\begin{figure}[t]
\setlength{\unitlength}{1mm}
\leavevmode
\hskip  0mm
\includegraphics[width=170mm,bb=45 290 510 575,clip]{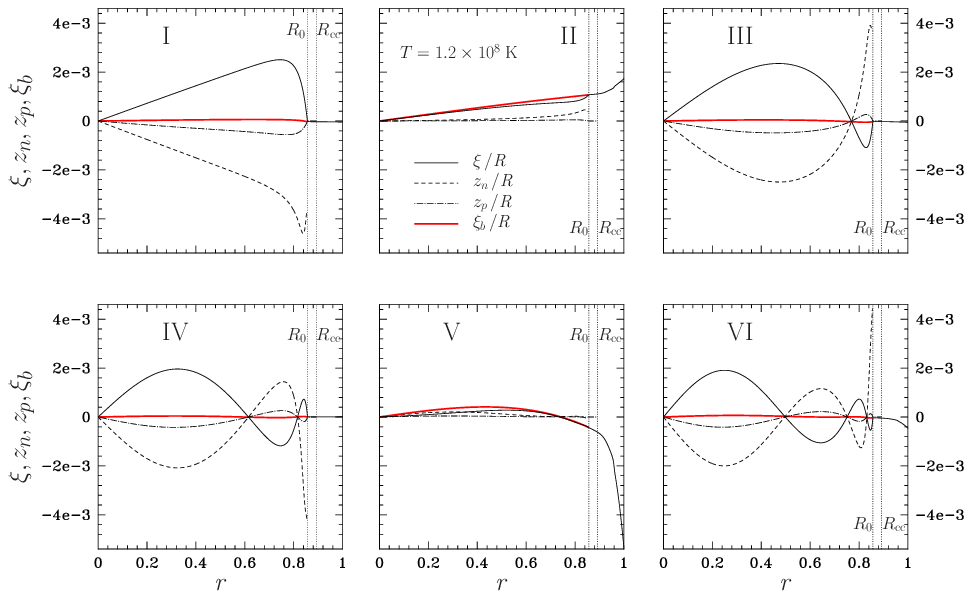}
\caption{
Eigenfunctions $\xi$, $z_{ n}$, $z_{ p}$, and $\xi_{ b}$ 
(normalized so that $E_{\rm puls}=10^{48}\, \rm erg$) 
of a superfluid NS at $T^\infty=1.2\cdot 10^8 \, \rm K$ 
versus $r$ (in units of $R$), calculated 
for the first six pulsation modes (I,...,VI) 
of the model 3. 
Solid curves correspond to $\xi$, 
dashed -- to $z_{ n}$, dot-dashed -- to $z_{ p}$, 
heavy solid lines show $\xi_{ b}$. 
Vertical dots indicate the crust-core boundary 
and the superfluid-normal interface.
}
\label{fig3}
\end{figure}

At the chosen temperature 
($T^\infty=1.2\times 10^8 \,\rm K$), 
the modes I, III, IV, and VI experience oscillations of the superfluid type 
and are approximately described by Eq.\ (\ref{sflapp}). 
The modes II and V are of normal type
and are approximately described by Eq.\ (\ref{normapp}); 
their eignefrequencies roughly coincide 
with the first and second modes 
of a normal star of the same mass. 
It can be seen that, as expected, 
the baryon current in superfluid oscillations, 
characterized by the quantity $\xi_{ b}$, is almost zero. 

It is noteworthy that the crust of a star, 
experiencing superfluid oscillations, remains nearly at rest, 
i.e., pulsations are mostly localized in the superfluid region. 
This point is illustrated in Fig.\ 6, 
which presents the ratio of average oscillation energy densities 
in the nonsuperfluid $\varepsilon_{\rm norm}$ and superfluid $\varepsilon_{\rm sfl}$ 
regions as a function of $T^\infty$ 
for the first six pulsation modes (I,...,VI).
Modes I, II, and III are shown, respectively, by the
solid, dotted, and dashed lines in the left panel;
the other modes are shown in the right panel. 
The thin solid curve is plotted 
for the first two modes of a normal star 
(I$_{\rm nfh}$ and II$_{\rm nfh}$).
The latter modes depend on temperature 
only because of the variation of $R_0$ with $T^\infty$,
which is artificially assumed to be the same 
as for a superfluid star.
 
It can be seen from Fig.\ 6 that, 
for normal-type oscillations, 
the ratio $\varepsilon_{\rm norm}/\varepsilon_{\rm sfl}$
coincides very well with the result for normal star.
However, immediately after the star starts 
to experience pulsations of the superfluid type, 
$\varepsilon_{\rm norm}/\varepsilon_{\rm sfl}$
falls nearly to zero, 
so that almost all pulsation energy
will be localized within the superfluid region. 
Thus, we see that the superfluid pulsation modes 
practically do not appear on the stellar surface and 
therefore it should be difficult to observe them
by detecting the modulation of electromagnetic 
radiation from the star.

\begin{figure}[t]
\setlength{\unitlength}{1mm}
\leavevmode
\hskip  0mm
\includegraphics[width=150mm,bb=10 290 560 560,clip]{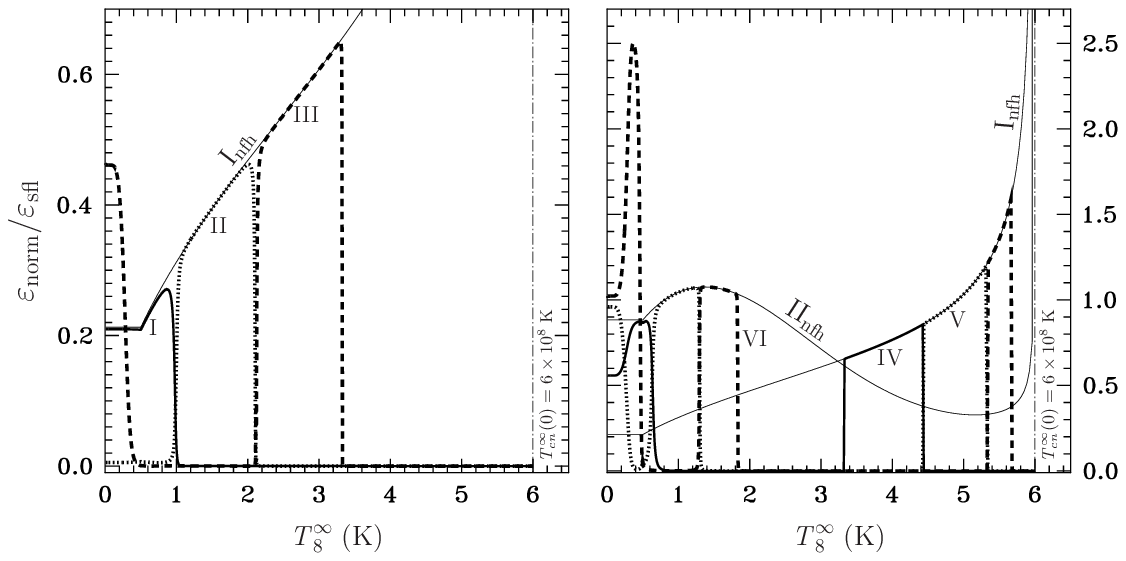}
\caption{
The ratio
of the pulsation energy densities 
in the nonsuperfluid and superfluid regions 
of a star versus $T^\infty_8$, 
calculated for model 3 (see the text). 
The left panel shows the first three modes (I, II, and III), 
the right panel shows the modes IV, V, and VI. 
Solid thin lines 
present the same ratio 
for the first two pulsation modes
of a normal star (I$_{\rm nfh}$ and II$_{\rm nfh}$). 
The redshifted neutron critical temperature 
$T_{c n}^\infty$ in the center of the star
is indicated by the vertical dot-dashed lines. 
}
\label{energy}
\end{figure}

\section{Damping times}

In this section
we calculate characteristic damping times 
for radial pulsations 
due to the shear viscosity 
and nonequilibrium reactions 
of mutual particle transformations. 
It is well known that such reactions 
generate an effective bulk viscosity 
(see, e.g., \cite{dsw07,gusakov07} and references therein).

The e-folding time is defined as
\begin{equation}
\tau = - \frac{2E_{\rm puls}}{\langle \dot{E}_{\rm puls} \rangle}, \label{tau}
\end{equation}
where 
$\langle\dot{E}_{\rm puls}\rangle$ 
is the damping rate for $E_{\rm puls}$, 
averaged over the period $2 \pi/\omega$. 
Here and below, the angular brackets denote averaging over the pulsation period.

The expression for the pulsation energy 
of a superfluid symmetric nonrotating star has the form
\begin{eqnarray}
E_{\rm puls} = \frac{1}{2} \, \int \omega^2 
\left[ \mu_{i0} \mu_{k0} \, Y_{ik} \, z_{i\,(a)}  z_{k\,(a)} 
+ \mu_{i0} \mu_{k0} \, Y_{ik} \, \xi_{(a)} \, (z_{i\,(a)}+z_{k\,(a)}) + (P_0+\varepsilon_0) \, \xi_{(a)}^2\right]
\nonumber\\
\times  4 \pi r^2 \, {\rm e}^{3\lambda_0/2-\nu_0/2} \, dr,
\label{pulsation_energy}
\end{eqnarray}
where $z_{i\,(a)}$, $z_{k\,(a)}$, and $\xi_{(a)}$ 
are the amplitudes of $z_{i}$, $z_{k}$, and $\xi$, respectively. 
Using the symmetry of the matrix $Y_{ik}$, 
the quasineutrality condition (\ref{quasineutrality}), 
and the definition (\ref{xib}) of $\xi_{ b}$, 
we can simplify the expression for $E_{\rm puls}$,
\begin{eqnarray}
E_{\rm puls} = \frac{1}{2} \, \int \omega^2 
\left[ \mu_{ n0} \mu_{k0} \, Y_{{ n}k} \, z_{k\,(a)} \right( z_{{ n}\,(a)}-\frac{\mu_{i0} \, Y_{{ n}i} \, z_{i\,(a)}}{n_{ b0}}\left) + n_{ b0}\mu_{ n0} \, \xi_{{ b}\,(a)}^2\right] \nonumber \\
\times  4 \pi r^2 \, {\rm e}^{3\lambda_0/2-\nu_0/2} \, dr.
\label{energy2}
\end{eqnarray}
For a stellar model considered here, 
the direct Urca process is forbidden, 
and the only effective processes of particle transformations are the 
neutron and proton branches of the modified Urca process,
\begin{eqnarray}
&&n+n \rightarrow p+n+e^- + \overline{\nu}_e,\,\,\,\,\,\, p+n+e^- \rightarrow n+n+\nu_e; \label{murcan}\\
&&n+p \rightarrow p+p+e^- + \overline{\nu}_e,\,\,\,\,\,\, p+p+e^- \rightarrow n+p+\nu_e. \label{murcap}
\end{eqnarray}
These reactions lead to dissipation of $E_{\rm puls}$ 
and to heating of the star. 
Assuming that pulsations are subthermal ($\delta \mu \ll kT$), 
the conversion rate of the pulsation energy 
into the thermal energy (per unit volume)
via the nonequilibrium reactions 
(\ref{murcan}) and (\ref{murcap}), 
is given by \cite{kg09,ll87,gyg05}
\begin{equation}
Q_{\rm bulk}=\lambda \delta \mu^2,
\label{Qbulk}
\end{equation}
where the chemical potential imbalance 
$\delta \mu$ can be expressed in terms 
of the above-calculated Lagrangian displacements 
$\xi$, $z_{ n}$, and $z_{ p}$ 
with the help of Eqs.\ (\ref{expanddmu}), (\ref{dnb}), and (\ref{dne}); 
and $\lambda \equiv \lambda_1+\lambda_2$ 
is the sum of the reaction rates 
(\ref{murcan}) and (\ref{murcap}). 
The reaction rates and the corresponding reduction factors 
were calculated for the modified Urca process in Ref.\ \cite{hly01} 
(see also Ref.\ \cite{hly02}, 
in which a correction was introduced
to the reaction rate for the proton branch). 

The energy converted into heat per unit volume 
per second due to the shear viscosity 
is given by the same formula as for normal matter \cite{gyg05},
\begin{equation}
     {Q}_{\rm shear} = {\eta \over 3} \,\, \omega^2 \,\,
     { {\rm e}^{-\nu_0} \over r^2} \,\, \left(
     -2 r { \dd \xi \over \dd r}
     + 2 \xi  + r \xi
     {\dd \nu_0 \over \dd r} \right)^2,
\label{Qshear}
\end{equation}
where $\eta$ is the shear viscosity coefficient. 
In our numerical calculations we used $\eta$ from \cite{sy08}. 
Because the shear viscosity is mostly determined 
by the electron-electron and 
%
%
electron-proton collisions, 
we neglect the contribution of neutron collisions 
and use equation (37) from \cite{sy08} when calculating $\eta$. 
In addition, we use equation (83) from the same reference 
to take into account the reduction of the particle collision frequency 
(and, consequently, the increase in the shear viscosity) by the proton superfluidity.

To obtain the dissipation rate of the pulsation energy, 
one has to average Eqs.\ (\ref{Qbulk}) and (\ref{Qshear}) 
over the pulsation period and then
to integrate them over the stellar volume \cite{gyg05},
\begin{equation}
\langle \dot{E}_{\rm puls} \rangle=-\int (\langle Q_{\rm shear}\rangle
+\langle Q_{\rm bulk}\rangle){\rm e}^{\nu_0} 4 \pi r^2 {\rm e}^{\lambda_0/2} dr,
\label{EdotInt}
\end{equation}

Having thus expressed all the quantities in Eq.\ (\ref{tau}) 
in terms of the already calculated eigenfunctions, 
we can present numerical results for the e-folding times. 
As in Sec.\ V, we consider only the model 3.

\begin{figure}[t]
\setlength{\unitlength}{1mm}
\leavevmode
\hskip  0mm
\includegraphics[width=150mm,bb=15 295 570 560,clip]{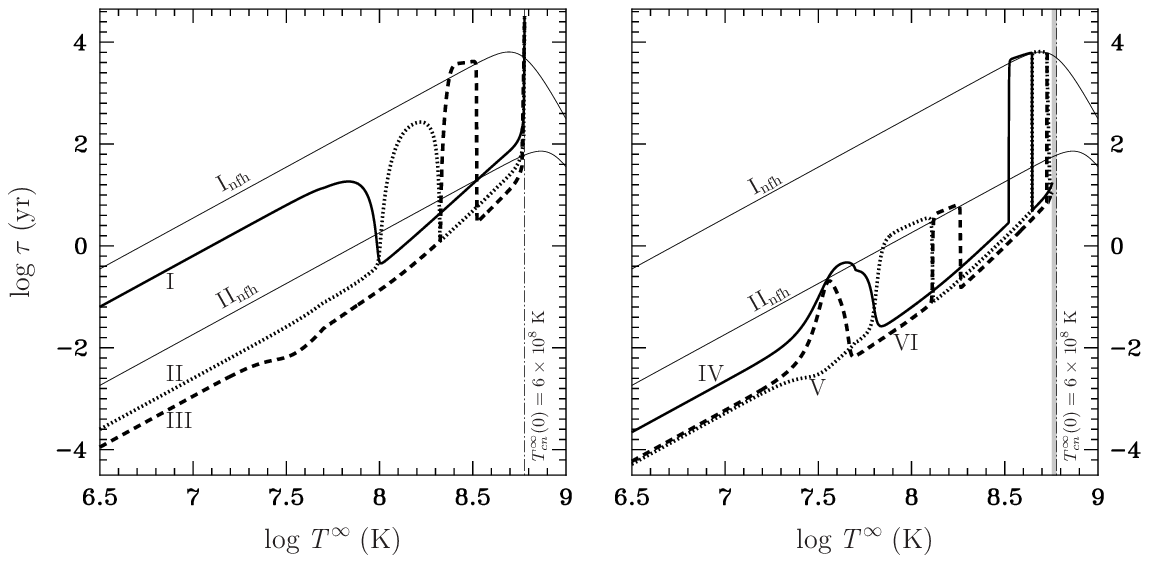}
\caption{
Damping time $\tau$ 
versus $T^{\infty}$
for the first six pulsation modes of model 3 (see the text). 
The thin solid lines
indicate 
$\tau$
for the first two modes of a normal star
(I$_{\rm nfh}$ and II$_{\rm nfh}$). 
Because of numerical problems $\tau$ was not calculated
in the filled region.
Other notations are the same as in Fig.\ 6.}
\label{tau1}
\end{figure}

Figure 7 presents $\tau$ as a function of $T^\infty$, 
for the first six pulsation modes (three modes in each panel).
As in Fig.\ 6, 
the modes are shown by heavy solid, dotted, and dashed lines. 
Thin solid lines show $\tau$ calculated 
for the first two modes I$_{\rm nfh}$ and II$_{\rm nfh}$ by using 
normal-fluid (ordinary) hydrodynamics.
In these calculations, 
superfluidity was taken into account 
only at calculating the kinetic coefficients 
(such an approximation is frequently employed in the literature 
to estimate the effect of superfluidity on the dissipative processes in NSs). 
The damping times were not calculated for the modes IV, V, and VI 
in the immediate vicinity of $T_{c  n}^\infty(0)$ 
because of problems with numerical calculations
(this region is filled grey).

It follows from Figs.\ 3 and 7 
that $\tau$ for normal-type oscillations may 
differ severalfold from that calculated 
using the normal-fluid hydrodynamics. 
The difference is stronger at lower temperatures. 
This fact may seem surprising if we recall that 
the eigenfrequencies and eigenfunctions $\xi_b(r)$ 
approximately coincide for normal modes 
and for corresponding modes
of a nonsuperfluid star.
To explain the fact, one should notice that
the function $\xi(r)$ 
generally differs in superfluid and normal stars.
In additon, 
the relative motion of superfluid components 
with respect to the normal one
($z_i \neq 0$)
also modifies $\tau$.

Superfluid pulsation modes are damped faster 
than the normal modes, so that 
$\tau$ drops by $1-3$ orders of magnitude 
each time the 
star 
switches (e.g., in the course of thermal evolution)
from normal- to superfluid-type regime of pulsations (see Figs.\ 3 and 7).  
To explain such a fast damping, 
let us note that it occurs 
(for $T^\infty \la 6 \times 10^8$ K) 
mainly due to the electron shear viscosity. 
For a nonsuperfluid matter 
the electron shear viscosity coefficient $\eta_e$
is smaller in the outer stellar layers than in
the inner layers. 
Proton superfluidity further increases $\eta_e$ 
in the inner layers of the star.
Now, if one looks at Fig.~\ref{fig3}, one will notice that
superfluid modes are localized in the superfluid core, 
while the amplitude of normal modes 
(modes II and V in that figure)
grows away from the stellar center 
(with the maximum in the crust, where $\eta_e$ is small).
In view of these facts 
the reason for the difference of damping times becomes clear;
quantitatively, this result follows from
a careful examination
of Eqs.\ (\ref{tau}), (\ref{Qshear}), and (\ref{EdotInt}).

In order to understand the limiting behavior of $\tau$ 
at high temperatures [$T^\infty \rightarrow T_{c  n}^\infty(0)$], 
let us analyze its asymptotes for different models. 
As already discussed in Sec.\ IV, 
any given mode near the neutron critical temperature 
experiences superfluid-type pulsations, 
approximately described by Eq.\ (\ref{sflapp}). 
For this type of pulsations $\xi_{ b}$ is small,
$\xi_{ b}\ll \mu_{k0} \, Y_{{ n}k} \, z_{k}/n_{ b}$, 
so that one derives from Eq.\ (\ref{energy2}) 
the following estimate 
for $E_{\rm puls}$ at $T^\infty \rightarrow T_{c  n}^\infty(0)$
\begin{equation}
E_{\rm puls} \sim \omega^2 \mu_{ n0} \mu_{k0} \, Y_{{ n}k} \, \widetilde{z}_{k\,(a)} \left[ \widetilde{z}_{{ n}\,(a)}-\frac{\mu_{i0} \, Y_{{ n}i} \, \widetilde{z}_{i\,(a)}}{n_{ b0}} \right] R_0^3. 
\label{estimate1}
\end{equation}
Here and below, we neglect the factors 
of the order of unity in all formulas 
and replace integration over the stellar volume with 
the factor $R_0^3$; 
the quantities $\widetilde{z}_{n\,(a)}$ and $\widetilde{z}_{p\,(a)}$ 
in Eq.\ (\ref{estimate1})
denote $z_{n\,(a)}$ and $z_{p\,(a)}$, 
averaged over the superfluid region. 
Taking into account that for superfluid modes
$\xi \approx -\mu_{k0} \, Y_{{ n}k} \, z_{k}/n_{ b0}$ 
one obtains, using Eqs.\ (\ref{Qbulk}) and (\ref{Qshear}), 
the following estimate for the 
dissipation rate 
due to, respectively, the shear viscosity 
and nonequilibrium modified Urca reactions (bulk viscosity)
\begin{eqnarray}
\langle \dot{E}_{\rm puls \,\, sh}\rangle 
\sim -\eta \omega^2 \left(\frac{\mu_{k0} 
\, Y_{{ n}k} \, \widetilde{z}_{k\,(a)} }{n_{ b0}} \right)^2 R_0,\\
\langle \dot{E}_{\rm puls \,\, b} \rangle 
\sim -\lambda n_{ e0}^2 \left(\frac{\mu_{k0} \, 
Y_{{ n}k} \, \widetilde{z}_{k\,(a)} }{n_{ b0}}\right)^2 
\left(\frac{\partial \delta \mu}{\partial n_{ e}}\right)^2 R_0.
\label{bulk2}
\end{eqnarray}
In view of Eqs.\ (\ref{estimate1})--(\ref{bulk2})
one finds for the corresponding e-folding times, 
\begin{eqnarray}
\tau_{\rm sh} \equiv -\frac{2\,E_{\rm puls}}{ \langle \dot{E}_{\rm puls\,\, sh} \rangle} 
\sim \frac{\mu_{ n0} \left[ \widetilde{z}_{{ n}\,(a)}-\mu_{i0} \, Y_{{ n}i} 
\, \widetilde{z}_{i\,(a)}/n_{ b0} \right] 
R_0^2 n_{ b0}^2}{\eta \mu_{k0} \, Y_{{ n}k} \, \widetilde{z}_{k\,(a)}},\\
\tau_{\rm b} \equiv -\frac{2\,E_{\rm puls}}{\langle \dot{E}_{\rm puls\,\, b} \rangle} 
\sim \frac{\omega^2 \mu_{ n0} \left[ \widetilde{z}_{{ n}\,(a)}-\mu_{i0} 
\, Y_{{ n}i} \, \widetilde{z}_{i\,(a)}/n_{ b0} \right] 
R_0^2 n_{ b0}^2}{\lambda n_{ e0}^2 \mu_{k0} \, Y_{{ n}k} 
\, \widetilde{z}_{k\,(a)} \left(\partial \delta \mu/\partial n_{ e}\right)^2}.
\end{eqnarray}
Now, using the asymptote (\ref{Yikasy}) for $Y_{ik}$ 
and the asymptotes for $\omega$ and $R_0$, 
derived in Sec.\ IV, one gets 
for $\tau_{\rm sh}$ and $\tau_{\rm b}$
in the limit $T^\infty \rightarrow T_{c  n}^\infty(0)$,
\begin{eqnarray}
\tau_{\rm sh} &\propto& \frac{R_0^2}{ Y_{{ nn}} } \rightarrow \left(1- \frac{T^{\infty}}{T_{cn}^\infty}\right)^{\alpha},\\
\tau_{\rm b} &\propto& \frac{\omega^2 R_0^2}{Y_{{ nn}}} \rightarrow \rm finite.
\end{eqnarray}
Here $\alpha=-1$ for the first model, 
$\alpha=1/2$ for the second, 
$\alpha=-1/2$ for the third, 
and $\alpha=0$ for the fourth model. 
The damping time $\tau_{\rm b}$ 
tends to some finite value
for all the models.
For the third model, presented in Fig.\ 7,
the damping rate due to the shear viscosity 
drops as $\sqrt{1-T^\infty/T_{c  n}^\infty}$ 
at $T^\infty \rightarrow T_{c  n}^\infty(0)$ 
(that is, $\tau_{\rm sh}$ increases).
Hence, very close to $T_{c  n}^\infty(0)$, 
$1/\tau_{\rm sh} \ll 1/\tau_{\rm b}$ 
so that $\tau=\tau_{\rm sh}\tau_{\rm b}/(\tau_{\rm sh}+\tau_{\rm b})$ 
is bounded from above by $\tau_{\rm b}$
(this region of $T^\infty$ 
is too narrow to be visible in the figure).

\section{Summary}

This work analyses the effects of finite temperatures 
on the oscillations of superfluid NSs.
The most simple case of a radially pulsating nonrotating 
star is considered in the frame of general relativity.
The impact of two temperature effects 
on the pulsations is examined.
The first effect concerns the temperature dependence 
of the relativistic entrainment matrix $Y_{ik}$ --
one of the important ingredients of superfluid 
hydrodynamics. 
The second effect consists in 
decreasing of the size 
of superfluid region with increasing temperature.
The first effect was studied in GA06, the second -- 
in Refs.\ \cite{lac08,hap09,ha10}. 

We considered four different models which 
allow for
one or both of these 
effects, and
employ various profiles $T_{cn}^\infty(r)$
of neutron critical temperature.
We calculated pulsation spectra 
for these models.
For each model the presence of 
two distinct classes of oscillation modes,
the so called normal and superfluid modes, is revealed.
The frequencies of normal modes
almost coincide with the corresponding frequencies 
of a nonsuperfluid star. 
On the contrary, 
the frequencies of superfluid modes strongly depend 
on temperature and show different type
of behavior 
depending on a model.
It is demonstrated that the model 
which takes into account 
only the second temperature effect
(dependence of the size of superfluid region on $T$) 
but ignores the first effect, 
yields {\it qualitatively incorrect} spectra.

Using the approach developed in Refs.\ \cite{gk10a,gk10b},
we found the asymptotes for eigenfrequencies
at $T^\infty \rightarrow T_{c  n}^\infty(0)$,
which 
are in a good agreement 
with the results of numerical calculations.

In addition, we analyzed the eigenfunctions 
for the first six oscillation modes. 
It is shown that the baryon current density
is almost unperturbed in superfluid modes. 
Moreover, these modes are mainly localized in the core 
and practically do not appear on the neutron-star surface. 
As a consequence, it might be difficult to detect them
by observing modulation 
of the electromagnetic radiation from the star.

We also calculated the characteristic damping 
times $\tau$ for radial pulsations 
due to the shear viscosity 
and nonequilibrium modified Urca processes 
and analyzed the asymptotes of $\tau$ 
at $T^\infty \rightarrow T_{c  n}^\infty(0)$. 
It is shown that damping times, calculated 
using the ordinary (nonsuperfluid) hydrodynamics 
may differ severalfold from  
$\tau$ calculated for normal modes 
(even though the eigenfrequencies of normal modes 
almost coincide 
with the corresponding eigenfrequencies 
of a nonsuperfluid star). 
It is also demonstrated that, generally,
damping of superfluid-type pulsations occurs
by 1--3 orders of magnitude faster than
damping of normal-type pulsations.

\section*{Acknowledgments}

We thank A.I. Chugunov
for valuable comments.
This study was supported by the Dynasty foundation, 
Ministry of Education and Science of Russian Federation 
(contract No. 11.G34.31.0001 
with SPbSPU and leading scientist G.G. Pavlov), 
RFBR 11-02-00253-a, and FASI (grant NSh-3769.2010.2).

\end{document}